\documentclass[aps,showpacs]{revtex4}
\usepackage{epsf,epsfig,graphicx}
\begin{document}
\title{Glauber gluons in pion-induced Drell-Yan processes}

\author{Chun-peng Chang$^{1,2}$}
\author{Hsiang-nan Li$^{1,2,3}$} \email{hnli@phys.sinica.edu.tw}

\affiliation{$^1$Institute of Physics, Academia Sinica, Taipei,
Taiwan 115, Republic of China,}

\affiliation{$^2$Department of Physics, National Tsing-Hua
University, Hsinchu, Taiwan 300, Republic of China}

\affiliation{$^3$Department of Physics, National Cheng-Kung
University, Tainan, Taiwan 701, Republic of China}

\begin{abstract}

We point out that the existence of Glauber gluons in the $k_T$
factorization theorem can account for the violation of the Lam-Tung
relation, namely, the anomalous lepton angular distribution observed
in pion-induced Drell-Yan processes. The emission of a final-state
parton, that balances the lepton-pair transverse momentum, causes the responsible
spin-transverse-momentum correlation in the Glauber-gluon background.
It is argued that the Glauber effect is significant in the pion due to
its unique role of being a Nambu-Goldstone boson and a $q\bar q$ bound
state simultaneously. This mechanism is compared to other resolutions
in the literature by means of vacuum effects and Boer-Mulders functions.
We propose to discriminate the above resolutions by measuring the
$p\bar p$ Drell-Yan process at GSI and J-PARC.

\end{abstract}


\maketitle

Though a Drell-Yan process has been considered as being fully
understood, it reveals a puzzling behavior since 80's. The angular
distribution of produced lepton pairs is expressed
in the Collins-Soper (CS) frame as
\begin{eqnarray}
\frac{1}{\sigma}\frac{d\sigma}{d\Omega}=\frac{3}{4\pi}\frac{1}{\lambda+3}
\left(1+\lambda c_\theta^2 + \mu s_{2\theta}c_\phi +
\nu\frac{s_\theta^2}{2}c_{2\phi}\right),
\end{eqnarray}
with the notations $c_\theta\equiv \cos\theta$ and $s_\theta\equiv
\sin\theta$. In the above expression $\theta$ is the angle between
one of the leptons and the $z$ axis, and $\phi$ is the angle between
the plane formed by one of the lepton momenta and the $z$ axis and
the plane formed by the colliding beam momenta.
The involved coefficients obey the Lam-Tung
relation $1-\lambda-2\nu=0$ \cite{LT78}, which holds under
perturbative corrections \cite{MO95,Qiu}, and
for the inclusion of parton transverse
momentum and soft gluon effects \cite{CB86}. This relation has
been experimentally verified in the $pp$ and $pd$ Drell-Yan
processes \cite{E866}. However, significant violation was observed in
pion-induced ones, which
was found to increases with the lepton-pair transverse momentum $q_T$
\cite{NA10-86,NA10,E615}.

Resolutions of the above violation include the vacuum effect
\cite{BNM93,BNM05} that causes the transverse-spin correlation
between colliding partons, and the Boer-Mulders (BM) functions
\cite{Boer99}, that introduce the spin-transverse-momentum
correlation of a parton. The vacuum effect is flavor-blind
\cite{BNM05}, so it demands more effort to
differentiate the $\pi p$ and $pp$ Drell-Yan processes.
The BM proposal can differentiate the pion from the proton, because
the involved anti-quark is a valence (sea) parton in the former (latter).
In this letter we shall propose a
resolution by means of Glauber gluons in the $k_T$ factorization
theorem \cite{CQ07,C07}, whose effect might be significant due to the
unique role of the pion as a Nambu-Goldstone (NG) boson  and a
$q\bar q$ bound state simultaneously \cite{NS08}. The emission of
a final-state parton, that balances
the lepton-pair transverse momentum, causes the required
spin-transverse-momentum correlation in the Glauber-gluon background.
A simple discrimination is to measure the $p\bar p$ Drell-Yan process,
which will exhibit strong violation of the Lam-Tung relation according
to BM, but will not according to the uniqueness of the pion.

It has been known that the $k_T$ factorization of complicated QCD
processes involving more than two hadrons, such as hadron hadroproduction 
at the transverse momentum $q_T$ \cite{CQ07,C07}, is broken by residual
infrared divergences from the Glauber region. It was then demonstrated
that these Glauber divergences can be factorized into a new universal
nonperturbative phase factor at low $q_T$, where the eikonal
approximation holds even in the Glauber region \cite{CL09}. It is
obvious that the $q_T$ spectra of $\lambda$, $\mu$ and $\nu$ meet the
criteria for introducing the Glauber phase: since the low-$q_T$ spectra
are concerned, the $k_T$ factorization is an appropriate theoretical
framework; a final-state parton is required to balance the lepton-pair
$q_T$, so at least three partons are involved; the Glauber gluons exist
and are factorizable at low $q_T$, leading to the Glauber phase factor.

\begin{figure}[tb]
\begin{center}
\includegraphics[scale=0.5]{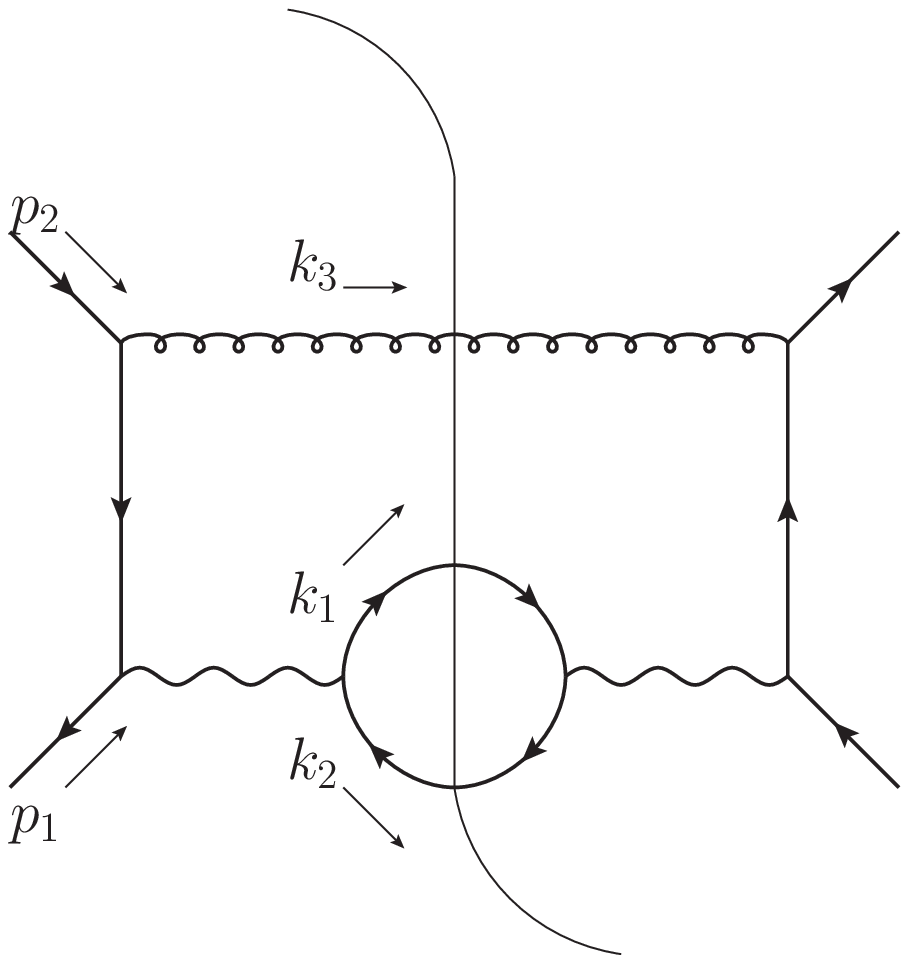}\hspace{1.0cm}
\includegraphics[scale=0.5]{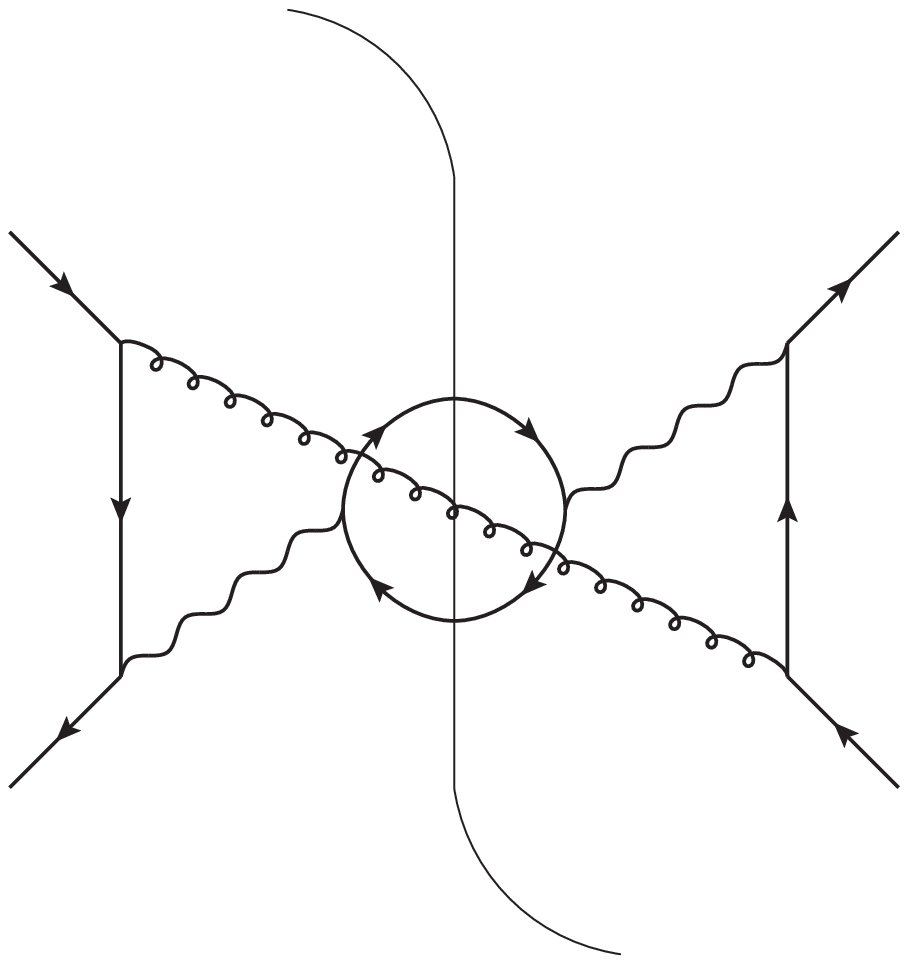}

(a) \hspace{5.5 cm} (b)

\caption{Some LO diagrams for lepton-pair production with $q_T$ in a
Drell-Yan process.}\label{fig1}
\end{center}
\end{figure}

The leading-order diagrams for the parton-level scattering
$\bar q(p_1)+q(p_2)\to \ell^-(k_1)+\ell^+(k_2)+g(k_3)$
in the pion-proton Drell-Yan process are displayed in Fig.~\ref{fig1}.
The diagrams with the gluon of momentum $k_3$ and the lepton pair being
exchanged is implicit. The anti-quark (quark) carrying the momentum
$p_1$ ($p_2$) is a parton of the pion (proton). The above momenta
are chosen, in the CS frame, as
\begin{eqnarray}
p_1 &=& E_1( 1, -s_1c_\phi, -s_1s_\phi, c_1), \nonumber\\
p_2 &=& E_2( 1, -s_1c_\phi, -s_1s_\phi, -c_1 ), \nonumber\\
k_1 &=& k(1, -s_\theta, 0, c_\theta),\;\;
k_2 = k(1, s_\theta, 0, -c_\theta),
\end{eqnarray}
and $k_3 = p_1 + p_2 - k_1 - k_2$, with $E_1$, $E_2$, and $k$ being the
energies, $\theta_1$ being the angle between the momentum $\bf p_1$ and
the $z$ axis, and the notations $c_1\equiv \cos\theta_1$ and
$s_1\equiv\sin\theta_1$. Note that $p_1$ and $p_2$ are slightly off
shell with $p_1^2<0$ and $p_2^2<0$ in the $k_T$ factorization. This
virtuality will not be shown explicitly, because it is not crucial for
the discussion below. $p_1$ ($p_2$) is mainly along the plus (minus)
direction, as $s_1$ is small. The analysis of another set of parton
scattering $gq\to \ell^-\ell^+ q$ is similar, which gives a minor
contribution in the region with finite parton momentum fractions.

\begin{figure}[tb]
\begin{center}
\includegraphics[scale=0.45]{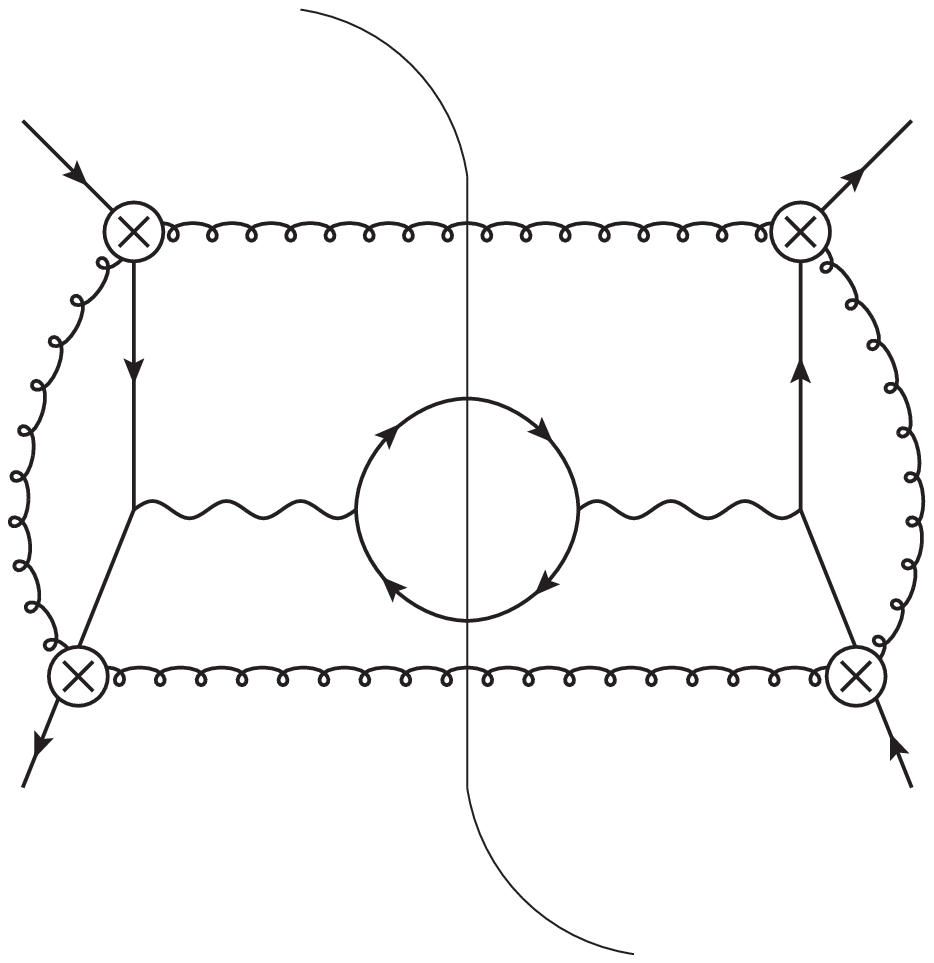}\hspace{1.0cm}
\includegraphics[scale=0.45]{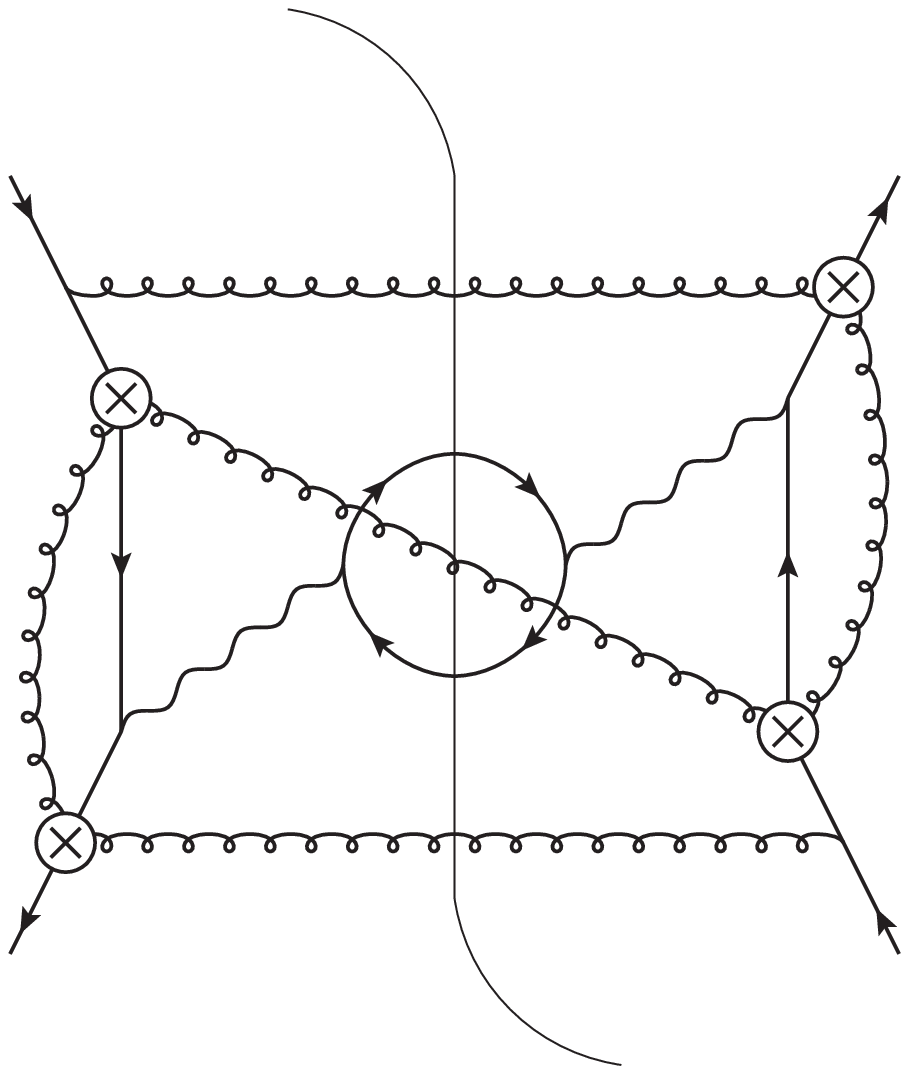}

(a) \hspace{5.5 cm} (b)

\caption{Glauber gluons (vertical lines) for each diagram in Fig.~\ref{fig1}.}\label{fig2}
\end{center}
\end{figure}

The Glauber divergence arises from a higher-order correction, where a
loop gluon of momentum $l$ connects, for example, a rung gluon
exchanged between the two partons in the pion, and the lines in the
upper half of Fig.~\ref{fig1}(a). The attachments to the parton in the
proton, the real gluon of momentum $k_3$, and the virtual quark
produce an imaginary piece $-2\pi i\delta(l^+)$ \cite{CQ07,C07}, with
$l^+$ being the plus component of $l$. The valence anti-quark propagator
and the rung gluon propagator on the pion side provide poles in the
different half complex planes of the minus component $l^-$ \cite{CQ07,C07}.
Therefore, the loop gluon is restricted in the Glauber region with tiny
virtuality $-l_T^2$, leading to the logarithmic Glauber divergence. The
factorization of a Glauber gluon can be achieved at low $q_T$ by applying
the eikonal approximation to the lines it connects, and by applying the
Ward identity to all the attachments \cite{CL09}. The outcome is displayed
in Fig.~\ref{fig2}(a), where the upper and lower vertices denoted by
$\otimes$ represent $\delta(l^+)$ and $\delta(l^-)$, respectively:
$\delta(l^-)$ gives a result the same as the contour
integration in the $l^-$ plane. Repeating the above procedures, we
factorize the Glauber gluons in Fig.~\ref{fig2}(b). Comparing
Fig.~\ref{fig2}(b) with \ref{fig2}(a), the existence of the Glauber
divergence in the former demands one more rung gluon. That is, it is
down by a power of $\alpha_s$, and regarded as a next-to-leading
logarithm. Exponentiating the imaginary logarithms in Fig.~\ref{fig2},
we obtain the Glauber phase factor $\exp(iS_e)$ \cite{CL09}.

The uniqueness of the pion is hinted by numerous anomalous data involving
pions in addition to the deviation of the Lam-Tung relation: the observed
$B^0\to\pi^0\pi^0$, $\pi^0\rho^0$, and $\rho^0\rho^0$ branching ratios are
6 times of, 3 times of, and consistent with the perturbative QCD (PQCD)
predictions \cite{LM06}. All these decays are dominated by the
color-suppressed tree amplitude $C$, and it is not understood why the
deviation becomes more significant as number of pions increases. This is
the so-called $B\to\pi\pi$ puzzle. The direct CP asymmetry in the
$B^+\to\pi^0 K^+$ decay is expected to be of the same sign as in
$B^0\to\pi^- K^+$, but the data indicate that they have opposite
signs. The former also involves $C$, so
the $B\to\pi K$ puzzle might have the same origin as of the $B\to\pi\pi$
one \cite{LMS05}. The data of the $D^0\to\pi^+\pi^-$ and $K^+K^+$ branching
ratios exhibit a difference much larger than expected in the
topological amplitude parametrization, even after the SU(3) symmetry
breaking has been included \cite{diag}. The $q_T$
spectrum of the pion hadroproduction is dramatically distinct from the
hadroproduction of other hadrons: fitting to the data from $pp$ collisions
at RHIC and LHC, and from $AuAu$ collisions with different centralities at RHIC
implies that the former follows an exponential law, but the latter follows a
power law \cite{BR12}. Nevertheless, the power-law contributions become dominant
in the pion production in the $\gamma p$ and $\gamma\gamma$ collisions
\cite{BR12}. We point out that all the anomalous processes demand the
$k_T$ factorization because of the end-point singularity in heavy-quark
decays, and the considered low-$q_T$ spectra. All the anomalous
processes involve at least three hadrons (note that the pion production
in the $\gamma p$ and $\gamma\gamma$ collisions involves fewer hadrons).
These common features are the necessary conditions for the Glauber gluons to
appear. It is certainly possible that the above puzzles have
different origins, but it would be very intriguing if they can be
resolved by the same mechanism.

The multiple Fock states of a pion have been proposed to reconcile its
simultaneous roles as a $q\bar q$ bound state and a NG boson \cite{NS08}.
It was then speculated that the Glauber effect becomes significant due
to the huge soft cloud formed by higher Fock states \cite{LM11}. Based on
this speculation, the constant phase factor $\exp(iS_e)$ was assigned
to the spectator amplitudes in the decays $B\to \pi\pi$,
$B\to\pi\rho$, $B\to\pi K$, and $D\to\pi\pi$. It was shown that a Glauber
phase in the range $S_e\sim (-\pi/4,-\pi/2)$ could resolve the puzzles
to some extent \cite{LM11,LLY12}. In this case it is
$\sin S_e$ from $\exp(iS_e)$ that is relevant to the interference
among various amplitudes, so the sign of $S_e$ matters.
The Glauber factor turns the destructive interference between the two
spectator diagrams into constructive one in the PQCD approach, such that
the color-suppressed tree amplitudes in the $B$ meson decays and the
$W$-exchange and $W$-annihilation amplitudes in $D$ meson decays are enhanced.
The enhanced amplitude $C$, modifying the interference
between the tree and penguin contributions to $B^+\to\pi^0 K^+$,
flips the sign of its direct CP asymmetry \cite{LM11}. The enlarged
$C$ also increases the $B^0\to\pi^0\rho^0$ and
$\pi^0\pi^0$ branching ratios by three times, such that the former agrees
with the data, and the latter is closer to, but still lower than the data
\cite{LM11}. The enhanced $W$-exchange and $W$-annihilation amplitudes
modify their interference with the emission diagrams in the way that the
$D^0\to\pi^+\pi^-$ branching ratio decreases \cite{LLY12}, and becomes
consistent with the data.

Putting the above clues together, we estimate the Glauber effect in the
pion-proton Drell-Yan process. It is the real part of
$\exp(iS_e)$, i.e., $c_{e}\equiv \cos S_e$, that contributes to a cross
section, so the sign of $S_e$ does not matter. Strictly speaking, the
Glauber phase factor appears in a $k_T$-convolution with the parton-level
hard scattering and the transverse-momentum-dependent parton distribution
functions of the pion and proton \cite{CL09}. Moreover, this phase should
vanish in the $q_T\to 0$ limit, where radiations are mainly soft and have
been resummed \cite{CSS88}, so the process involves only two
colliding hadrons. Hence, the constant phase
$S_e$ can be regarded as an average effect from the integration over the
parton transverse momentum. Assigning $c_e$ to Fig.~\ref{fig2}(a),
and neglecting the subleading Glauber effect in
Fig.~\ref{fig2}(c), we derive the parton-level hard kernel $\hat\sigma_i$
for each angular structure
\begin{eqnarray}
\hat\sigma_0&=&
\left(\frac{E_1}{E_2}+\frac{E_2}{E_1}\right)\left(1+\frac{1}{2}s_1^2\right)+(c_{e}-1)\nonumber\\
& &\times\left\{2\left[\frac{E_1E_2}{k^2}c_1^2+\frac{k}{E_1}
+\frac{k}{E_2}-\frac{1}{2}\left(\frac{E_1}{E_2}+\frac{E_2}{E_1}\right)
-2 \right]\right.\nonumber\\
& &\left.\;\;\;\;-\left[\frac{E_1E_2}{k^2}c_1^2
-\frac{1}{2}\left(\frac{E_1}{E_2}+\frac{E_2}{E_1}\right)\right]s_1^2\right\},\\
\hat\sigma_1&=&\left(\frac{E_1}{E_2}+\frac{E_2}{E_1}\right)\left(c_1^2-\frac{1}{2}s_1^2
\right)+ (c_{e}-1)\nonumber\\
& &\times\left\{\left(\frac{E_1}{E_2}
+\frac{E_2}{E_1}-2\right)c_1^2+\left[\frac{E_1E_2}{k^2}c_1^2
-\frac{1}{2}\left(\frac{E_1}{E_2}+\frac{E_2}{E_1}\right)\right]s_1^2\right\},
\\
\hat\sigma_2 &=&\left(\frac{E_1}{E_2}
-\frac{E_2}{E_1}\right)c_1s_1 +(c_{e}-1)
\left(\frac{E_2-E_1}{k}+\frac{E_1}{E_2}-\frac{E_2}{E_1}\right)c_1s_1,\\
\hat\sigma_3&= &\left(\frac{E_1}{E_2}
+\frac{E_2}{E_1}\right)s_1^2-(c_{e}-1)
2\left[\frac{E_1E_2}{k^2}c_1^2-\frac{1}{2}\left(\frac{E_1}{E_2}
+\frac{E_2}{E_1}\right)\right]s_1^2.
\end{eqnarray}
The terms proportional to $c_e-1$ are responsible for the anomalous
lepton angular diatribution.

The kinematic variables $E_1$, $E_2$, $k$, $s_1$ in the CS frame are then
transformed into $x_1P_1$, $x_2P_2$, $Q$, and $q_T$ in the center-of-mass
frame of the colliding hadrons, where $P_1$ and $P_2$ are the momenta of
the pion and the proton, respectively, $Q$ is the lepton-pair invariant
mass. Including the condition that the lepton pairs were produced mainly
at central pseudorapidity $\eta=0$, we have
\begin{eqnarray}
k=\frac{Q}{2},\;\;s_1^2=\frac{q_T^2}{Q^2+q_T^2},\;\;
E_1=cx_1P_1^0,\;\; E_2=cx_2P_2^0,
\end{eqnarray}
with the coefficient $c=\sqrt{(Q^2+q_T^2)/Q^2}$, and $P_1^0=P_2^0\equiv \sqrt{s}/2$
being the beam energy. It is seen that $s_1$ is proportional to $q_T$, i.e., to
the boost of the CS frame relative to the center-of-mass frame. The on-shell
condition $k_3^2=0$ relates the momentum fractions,
\begin{eqnarray}
x_2=\frac{x_1\sqrt{Q^2+q_T^2}-Q^2/\sqrt{s}}{x_1\sqrt{s}-\sqrt{Q^2+q_T^2}}.
\label{e12}
\end{eqnarray}
The requirement $0 \le x_2 \le 1$ sets the lower bound of $x_1$,
$x_{\rm min}=(\sqrt{Q^2+q_T^2}-Q^2/\sqrt{s})/
(\sqrt{s}-\sqrt{Q^2+q_T^2})$.

\begin{figure}[tb]
\begin{center}
\includegraphics[scale=0.6]{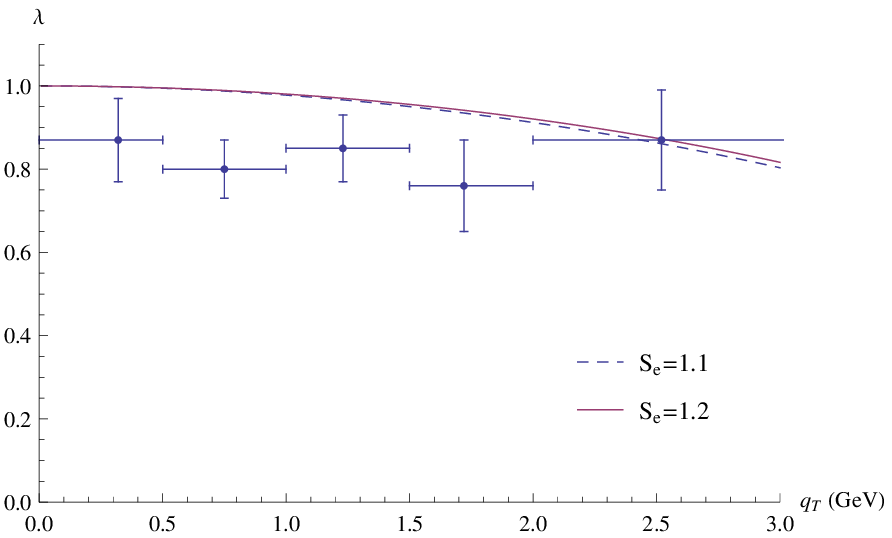}\hspace{0.5cm}
\includegraphics[scale=0.6]{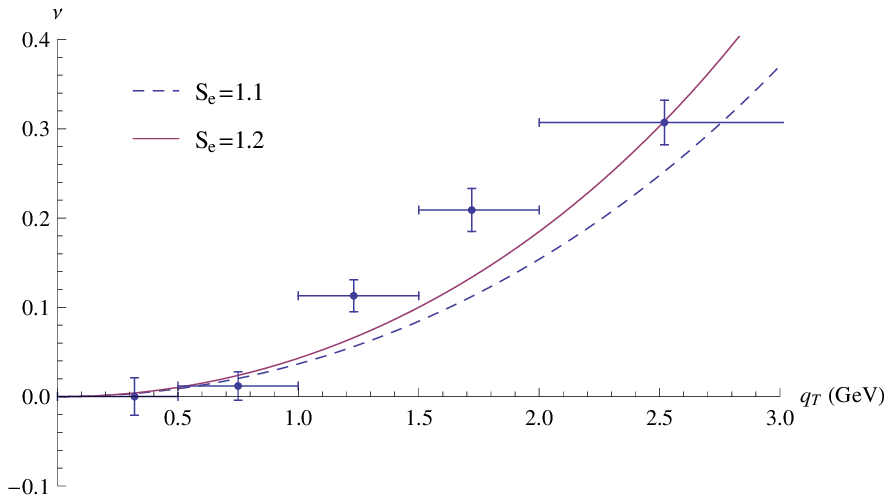}\hspace{0.5cm}
\includegraphics[scale=0.6]{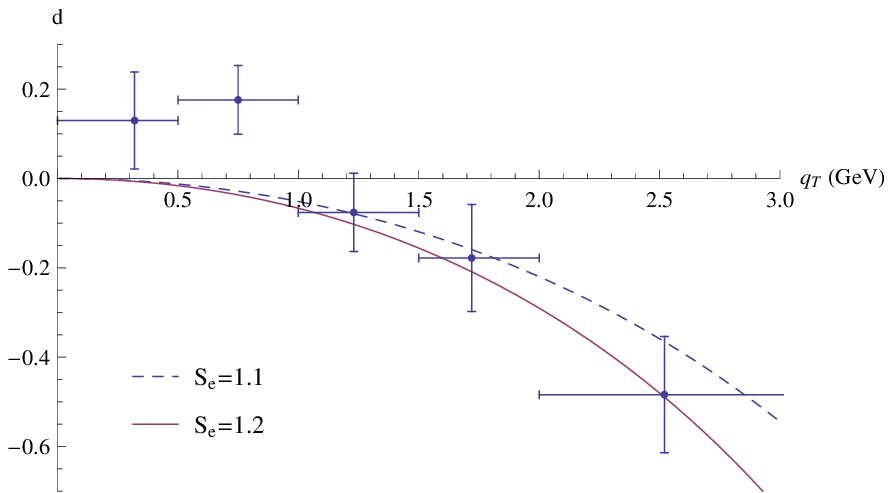}
\caption{$q_T$ dependence of $\lambda$, $\nu$ and the deviation $1-\lambda-2\nu$
for $S_e=1.1$ (dashed lines) and $S_e=1.2$ (solid lines).}\label{fig3}
\end{center}
\end{figure}

Convoluting the hard kernels with the pion (proton) parton distribution function
(PDF) $\phi_{\pi(p)}$, we obtain $\lambda=\sigma_1/\sigma_0$,
$\mu=\sigma_2/\sigma_0$, and $\nu=\sigma_3/\sigma_0$ with the definitions
\begin{eqnarray}
\sigma_i\propto \int_{x_{\rm min}}^1 \frac{\phi_\pi(x_1)\hat\sigma(x_1)\phi_p(x_2(x_1))}
{x_1\sqrt{s}-\sqrt{Q^2+q_T^2}}dx_1,
\end{eqnarray}
where the denominator comes from the Jacobian for the integration of
$\delta(k_3^2)$ over $x_2$. It is trivial to confirm that the Lam-Tung relation is
respected in the limit $S_e\to 0$. We choose the typical values $\sqrt{s}=194$ GeV
and $Q=8$ GeV \cite{BNM93}, and the PDFs $\phi_\pi(x)=\phi_p(x)\propto
x(1-x)^2$ for estimation. The Glauber phases $S_e=1.1$ and 1.2
lead to the predictions for $\lambda$, $\nu$, and the deviation $1-\lambda-2\nu$
in Fig.~\ref{fig3}, which agree well with the data \cite{NA10}. $\lambda$ is close to
unity and stable with respect to the variation of $S_e$ and $q_T$, and $\nu$
grows from zero with $q_T$ quickly. Note that $\nu$ and the deviation, being
proportional to $s_1^2$, diminish at $q_T=0$ naturally, where the Glauber
effect is supposed to disappear. $\mu$ is
equal to zero, when the pion and proton PDFs have exactly the same functional form,
and not presented here. That is, a nonvanishing $\mu$ can serve as a measurement of
the difference between the pion and proton PDFs. It is stressed that our
predictions for $\lambda$ and $\nu$ are insensitive to the choices of PDFs.

The discrimination between the resolutions based on the vacuum effect and
the BM functions by means of different $q_T$ and flavor dependencies
has been reviewed in \cite{BNM05}. Compared to the former, our proposal
also involves the breakdown of the standard $k_T$ factorization theorem,
which is, however, due to the nonperturbative mechanism
from the soft cloud of the pion: the Glauber gluons correlate the
quark and anti-quark distributions in hadron collision. The distinction
is that our resolution is flavor dependent, and able to differentiate
the $pp$ and $pd$ Drell-Yan processes from the pion-induced ones. In the
BM proposal the Lam-Tung relation holds in the $pp$ and $pd$ processes
because of the small sea-quark BM functions \cite{LS10}. Ours has
nothing to do with the valence or sea type of partons, so the two resolutions
can be discriminated by measuring the the $p\bar p$ Drell-Yan process. Since
the anti-quark in $\bar p$ is a valence parton, the associated BM
function is not suppressed. Then one should observe the violation of
the Lam-Tung relation, similar to what was observed in the pion-induced
Drell-Yan processes. According to our proposal, $\bar p$ is not a NG
boson, so the Glauber phase is not significant, and the Lam-Tung relation
should be respected. 

It can be shown that the Glauber divergence is power 
suppressed as one of the scales $Q$ and $q_T$ is large.
This power suppression is easily explained by means of Eq.~(10) in
\cite{C07}, the loop integral for the diagrams with one Glauber gluon on each side 
of the final-state cut, and Eq.~(11), the loop integral for the diagrams with two 
Glauber gluons on the same side of the cut. The similar loop integrals 
are derived for the above two types of diagrams in the present Drell-Yan analysis. However,
the rung gluon on the pion side, corresponding to the spectator particle in \cite{C07}, 
is a radiative gluon, so its transverse momentum $k_T$ should be integrated out. 
Applying the variable change $k_T - l_{2T} \to k_T$ to Eq.~(11) in \cite{C07}, where
$l_{2T}$ is the transverse momentum carried by the second Glauber gluon,
we see that Eq.~(10) and Eq.~(11) are almost identical but with a minus sign. 
The only difference is that $l_{2T}$ flows through the final-state parton
for Eq.~(11) after the variable change, but does not for Eq.~(10). If $Q$ is large,
the whole transverse components of the final-state parton 
momentum are negligible, and Eq.~(10) cancels Eq.~(11) exactly. If $q_T$
is large, $l_{2T}$ is negligible relative to $q_T$. The final-state parton
then carries the same momentum in Eq.~(10) and Eq.~(11), and their Glauber divergences 
also cancel each other exactly. This observation\footnote{We acknowledge J.C. Collins 
for pointing out this power suppression.} agrees with what was obtained in 
\cite{Cbook}. Hence, we claim, the same as BM, that the Lam-Tung relation holds in the high 
$q_T$ region. The data \cite{NA10} considered here came from the region,
where neither $Q$ (4.05 GeV $< Q<$ 8.5 GeV) nor $q_T$ ($q_T <$ 3 GeV) is high.
That is, the Glauber gluon effect may be relevant to these data. 


In this letter we have demonstrated that the Glauber phase, close to those
adopted for resolving the puzzles in the heavy-quark decays, resolves the
violation of the Lam-Tung relation in the pion-induced Drell-Yan processes.
The measured low-$q_T$ spectra of $\lambda$, $\nu$ and the deviation
$1-\lambda-2\nu$ were accommodated with this single parameter in our
formalism. We have proposed to discriminate the different resolutions by measuring
the $p\bar p$ Drell-Yan process at low $Q$ and $q_T$, which can be done at GSI and J-PARC.
If violation is (not) observed, our (BM) proposal is false. It will be an
important measurement for exploring the internal structures of hadrons and
for understanding the correlation of colliding partons in Drell-Yan processes.
It is mentioned that the lepton angular distribution in the $p\bar p$ Drell-Yan process 
has been measured by CDF at the $Z$ pole, i.e., large $Q$, \cite{CDF}, and that the 
data are consistent with the Lam-Tung relation in a wide range of $q_T$, from few up to 
around 80 GeV. The kaon-induced Drell-Yan processes might not serve the purpose, because, 
on one hand, the kaon BM function is less constrained, and on the other hand,
the NG nature of the kaon is not as solid as that of the pion due to SU(3) symmetry
breaking. Whether the anomalous pion hadroproduction spectra is also
attributed to the NG nature of the pion will be investigated elsewhere.

\acknowledgments{
We thank D. Boer, S. Brodsky, W.C. Chang, J.C. Collins, T. Hatsuda,
C.S. Lam, T. Onogi, J.W. Qiu, and T.M Yan
for useful discussions. This work was supported by the National
Science Council of R.O.C. under the Grant No.
NSC-101-2112-M-001-006-MY3.}

\end{document}